\documentclass[letter,twocolumn]{jpsj3}
\usepackage{txfonts}
\usepackage{color,ulem}

\title{
Spin Resonance in the New-Structure-Type Iron-Based Superconductor CaKFe$_4$As$_4$
}

\author{
Kazuki~Iida$^1$\thanks{k\_iida@cross.or.jp}, 
Motoyuki~Ishikado$^1$, 
Yuki~Nagai$^{2,3}$, 
Hiroyuki~Yoshida$^4$, 
Andrew~D.~Christianson$^5$, 
Naoki~Murai$^6$, 
Kenji~Kawashima$^{7,8}$, 
Yoshiyuki~Yoshida$^8$, 
Hiroshi~Eisaki$^8$, 
and Akira~Iyo$^8$
}

\inst{
$^1$Neutron Science and Technology Center, Comprehensive Research Organization for Science and Society (CROSS), Tokai, Ibaraki 319-1106, Japan\\
$^2$CCSE, Japan  Atomic Energy Agency, Wakashiba, Kashiwa, Chiba 277-0871, Japan\\
$^3$Department of Physics, Massachusetts Institute of Technology, Cambridge, Massachusetts  02139, USA\\
$^4$Department of Physics, Faculty of Science, Hokkaido University, Sapporo, Hokkaido 060-0810, Japan\\
$^5$Quantum Condensed Matter Science Division, Oak Ridge National Laboratory, Oak Ridge, Tennessee 37831, USA\\
$^6$J-PARC Center, Japan Atomic Energy Agency (JAEA), Tokai, Ibaraki 319-1195, Japan\\
$^7$IMRA Material R\&D Co., Ltd., Kariya, Aichi 448-0032, Japan\\
$^8$National Institute of Advanced Industrial Science and Technology (AIST), Tsukuba, Ibaraki 305-8568, Japan\\
}

\abst{
The dynamical spin susceptibility in the new-structure-type iron-based superconductor CaKFe$_4$As$_4$ was investigated by using a combination of inelastic neutron scattering (INS) measurements and random phase approximation (RPA) calculations.
Powder INS measurements show that the spin resonance at $Q_\text{res}=1.17(1)$~$\text{\AA}^{-1}$, corresponding to the $(\pi,\pi)$ nesting wave vector in tetragonal notation, evolves below $T_\text{c}$.
The characteristic energy of the spin resonance $E_\text{res}=12.5$~meV is smaller than twice the size of the superconducting gap ($2\Delta$).
The broad energy feature of the dynamical susceptibility of the spin resonance can be explained by the RPA calculations, in which the different superconducting gaps on different Fermi surfaces are taken into account.
Our INS and PRA studies demonstrate that the superconducting pairing nature in CaKFe$_4$As$_4$ is the $s_\pm$ symmetry.
}

\begin{document}
\maketitle

Various iron-based superconducting materials with different crystal structures have been investigated to transcend our understanding of the nature of their pairing mechanism~\cite{Review_1,Review_2,Review_3,Review_4}, as the crystal structure is believed to have a strong relationship with the superconducting transition temperature $T_\text{c}$.~\cite{LeePlot_1,PnictogenHeight}
Recently, new-structure-type iron-based superconducting materials $AeA$Fe$_4$As$_4$ (where $Ae$ = Ca, Sr, or Eu and $A$ = K, Rb, or Cs) were reported.~\cite{1144_powder,EuRbFe4As4_1,EuRbFe4As4_2}
The $AeA$Fe$_4$As$_4$ family exhibits $T_\text{c}=32$--37~K, which is relatively high compared with other stoichiometric iron-based superconductors.~\cite{FeSe,LiFeAs,KFe2As2,FeS,KCa2Fe4As4F2}
As $Ae$ and $A$ layers stack alternatively between Fe$_2$As$_2$ layers owing to the large difference between their ionic radii, $AeA$Fe$_4$As$_4$ belongs to the $\rm P4/mmm$ space group, which is different from $\rm I4/mmm$ in hole-doped (Ba$_{1-x}$K$_{x}$)Fe$_2$As$_2$ systems.~\cite{1144_powder,1144_single}
$AeA$Fe$_4$As$_4$ has high $T_\text{c}$ at  stochiometric composition and a unique crystal structure, providing us with a great opportunity to theoretically and experimentally investigate iron-based superconductivity in the absence of substitutional disorder.

CaKFe$_4$As$_4$ undergoes superconductivity below $T_\text{c}\simeq35$~K without other phase transitions for $1.8\le T\le300$~K.~\cite{1144_single,1144_Mossbauer,1144_pressure}
The magnetic ground state of the normal state in CaKFe$_4$As$_4$ is paramagnetic.~\cite{1144_Mossbauer,1144_uSR}
Various measurements~\cite{1144_single,1144_Mossbauer,1144_uSR,1144_Hc2,1144_penetration,1144_STM} indicated that CaKFe$_4$As$_4$ shows behaviors similar to optimally or slightly overdoped (Ba$_{1-x}$K$_x$)Fe$_2$As$_2$.
Angle-resolved photoemission spectroscopy (ARPES) measurements supported by density functional theory calculations indicated that the Fermi surface consists of three-hole pockets at the $\Gamma$ point and two electron pockets at the $M$ point.~\cite{1144_ARPES,CaKFe4As4_DFT}
This multiband nature implies multi-gap superconductivity in CaKFe$_4$As$_4$.
Two superconducting gaps which are nodeless and isotropic were, indeed, observed by scanning tunneling microscopy~\cite{1144_STM}, optical conductivity~\cite{1144_optical}, penetration depth~\cite{1144_penetration}, muon spin rotation~\cite{1144_uSR}, and $^{75}$As nuclear magnetic resonance (NMR)~\cite{1144_NMR} measurements.
The largest superconducting gaps ($\Delta_\text{hole}=13$~meV and $\Delta_\text{electron}=12$~meV) for quasi-two-dimensional hole and electron pockets that have similar diameters at the $\Gamma$ and $M$ points were reported by ARPES studies,~\cite{1144_ARPES} suggesting that the ideal nesting condition with the $(\pi, \pi)$ wave vector referred to the tetragonal reciprocal lattice.~\cite{CaKFe4As4_DFT,CaRbFe4As4_1}

These results indicated that CaKFe$_4$As$_4$ belongs to the same paradigm as other iron pnictide systems with respect to the presence of the disconnected Fermi surfaces that give rise to antiferromagnetic fluctuations in the wave vector $\mathbf{Q}_\text{AF}=(\pi, \pi)$.
One can, therefore, expect the same spin-fluctuation-mediated $s_\pm$-wave pairing mechanism~\cite{spm_1,spm_2} for this material, which was proposed in previous studies.~\cite{1144_STM,1144_penetration,1144_ARPES,CaKFe4As4_DFT,1144_NMR}
However, the sign of the superconducting gaps in CaKFe$_4$As$_4$ has never been investigated via a phase-sensitive technique.
As is well established, the appearance of  neutron spin resonance depends sensitively on the relative signs of the superconducting gaps on different portions of the Fermi surfaces separated by momentum $\mathbf{Q}_\text{AF}$, and thus it serves as a direct probe of the symmetry of the superconducting order parameter.~\cite{INStheory_1,INStheory_2}
For the $s_\pm$-pairing symmetry, the dynamical spin susceptibility ($\chi''$) is enhanced below $T_\text{c}$ by the Bardeen-Cooper-Schrieffer (BCS) coherence factors, and a well-defined resonance peak is formed. 
Therefore, investigation of the dynamical susceptibility to observe the neutron spin resonance is important for further understanding of the pairing symmetry in CaKFe$_4$As$_4$.
In this Letter, we report the dynamical spin susceptibility of CaKFe$_4$As$_4$ investigated by using inelastic neutron scattering (INS) measurements and random phase approximation (RPA) calculations.

\begin{figure}[t]
\includegraphics[width=8.47cm]{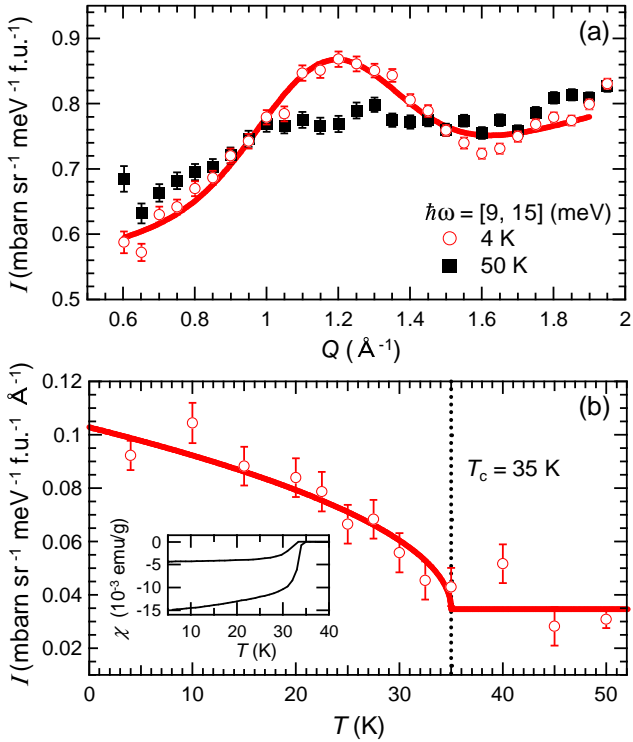}
\caption{\label{Fig:Q}
(Color online)
(a) $I(Q)$ in CaKFe$_4$As$_4$ at $T=4$ and 50~K with $E_\text{i}=70$~meV.
$\hbar\omega$ was integrated in $[9,15]$~meV.
The solid line represents the fitting result by using a Gaussian function with a linear background.
(b) Temperature dependence of the integral intensity of the spin resonance peak.
The intensity was estimated by fitting the Gaussian function with a linear background to the data at each temperature as shown in panel (a).
The solid line is a guide to the eye and the dashed line represents $T_\text{c}$ of the measured sample.
The inset shows the magnetic susceptibility of polycrystalline CaKFe$_4$As$_4$.
}
\end{figure}

Polycrystalline CaKFe$_4$As$_4$ was synthesized by following the procedure described in Ref.~[\citen{1144_powder}], and the susceptibility measurement shows $T_\text{c}=35$~K in the current sample [see the inset in Fig.~\ref{Fig:Q}(b)].
Powder CaKFe$_4$As$_4$ with a mass of $9.2$~g was used for our INS measurements.
Time-of-flight (TOF) neutron scattering measurements were performed using the wide-angular-range chopper spectrometer ARCS at the Oak Ridge National Laboratory (ORNL) Spallation Neutron Source (SNS) at a beam power of 1.2 MW~\cite{ARCS}.
Measurements with two different neutron incident energies $E_\text{i}=50$ and 70~meV were performed, and the energy resolutions at the elastic channel were 1.94 and 2.76~meV, respectively.
For the absolute scale in the present measurements, incoherent scattering from CaKFe$_4$As$_4$ was used.~\cite{AbsoluteNormalization}

\begin{figure}[t]
\includegraphics[width=8.47cm]{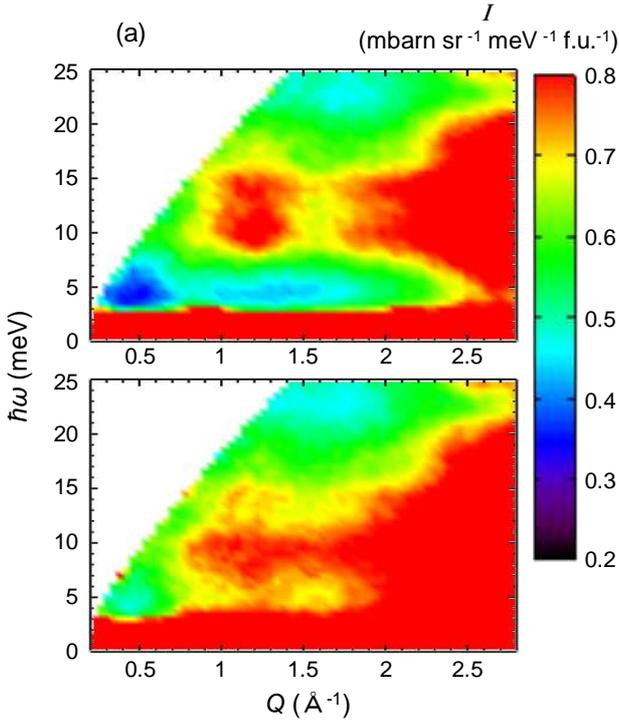}
\caption{\label{Fig:Map}
(Color online)
Neutron scattering intensity maps in CaKFe$_4$As$_4$ at (a) $T=4$~K and (b) 50~K with $E_\text{i}=50$~meV.
}
\end{figure}

Figures~\ref{Fig:Map}(a) and \ref{Fig:Map}(b) depict the neutron scattering intensity ($I$) maps from CaKFe$_4$As$_4$ at 4~K ($<$$T_\text{c}$) and 50~K ($>$$T_\text{c}$) as a function of momentum ($Q$) and energy $(\hbar\omega)$ transfers.
Below $T_\text{c}$, the spin resonance mode localized in both $Q$ $(\sim$$1.2~\text{\AA}^{-1})$ and $\hbar\omega$ $(\sim$$12~\text{meV})$ was observed as in other polycrystalline iron-based superconductors.~\cite{Ba122_1,1111_4,1111_2,111_1,Ba122_4,LiNH3Fe2Se2,1111_3,LiODFeSe_1}
The observed scattering intensity of the spin resonance in CaKFe$_4$As$_4$ is comparable with that of previous works.~\cite{1111_4,1111_2,111_1,Ba122_4,LiNH3Fe2Se2,1111_3,LiODFeSe_1}

\begin{figure}[t]
\includegraphics[width=8.47cm]{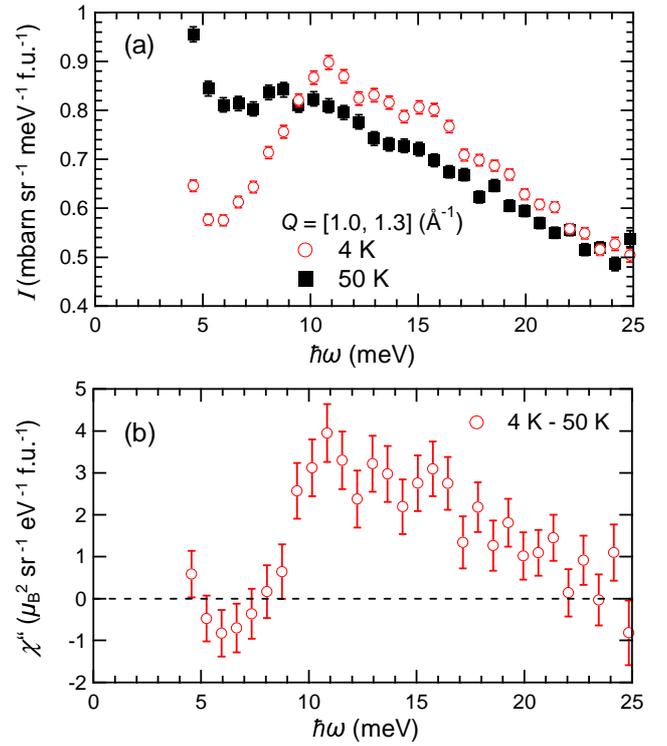}
\caption{\label{Fig:Energy}
(Color online)
(a) $I(\hbar\omega)$ in CaKFe$_4$As$_4$ at $T=4$ and 50~K.
(b) Temperature difference between the dynamical susceptibilities at 4 and 50~K.
In both panels, $E_\text{i}=70$~meV was used and $Q$ was integrated in $[1,1.3]$~$\text{\AA}^{-1}$.
}
\end{figure}

For quantitative analysis, energy cuts [$I(\hbar\omega)$] at $T=4$ and 50~K are plotted in Fig.~\ref{Fig:Energy}(a).
$Q$ was integrated in $[1,1.3]$~$\text{\AA}^{-1}$.
The intensity below $T_\text{c}$ was enhanced over the energy range of $10<\hbar\omega<25$~meV.
Compared with the spectrum at $T>T_\text{c}$, the spectral-weight gain in $10<\hbar\omega<25$~meV at 4~K is transferred by the depletion at lower energies, which is the typical behavior of the spin resonance in iron-based superconductors.~\cite{Ba122_1,1111_2}
To emphasize the neutron resonance mode, the temperature difference between the dynamical susceptibilities at 4 and 50~K after correction for the Bose factor and the squared magnetic form factor of Fe$^{2+}$~\cite{InternationalTalbes} was plotted in Fig.~\ref{Fig:Energy}(b).
The spectral weight is shifted to the spin resonance mode from the lower energy with a crossover energy of 8~meV.
The characteristic energy of the spin resonance is $\hbar\omega_\text{res}=12.5$~meV, which corresponds to $4.1(1)k_\text{B}T$ and $0.50(1)\times2\Delta$ (where $2\Delta=\Delta_\text{hole}+\Delta_\text{electron}$), in agreement with other iron-based superconductors.~\cite{Review_1,Review_2}
The observed $\chi''(\hbar\omega)$ spectrum of the spin resonance is broad.
As discussed later, our RPA calculation reproduces the broad spectral feature of the spin resonance anticipated as a result of the multiple superconducting gaps on the different Fermi surfaces in CaKFe$_4$As$_4$.

Cuts along $Q$ [$I(Q)$] at $T=4$ and 50~K by integrating $\hbar\omega$ in $[9,15]$~meV are also plotted in Fig.~\ref{Fig:Q}(a).
Although phonon scattering contributes to the scattering intensity at high $Q$ (see Fig.~\ref{Fig:Map}), the spin resonance is also sharp in $Q$ below $T_\text{c}$.
To quantify the peak position and the integral intensity of the spin resonance, a Gaussian function with a linear background was fitted to the $Q$ cut.
The peak center at 4~K is estimated to be $Q_\text{res}=1.17(1)$~$\text{\AA}^{-1}$, which is very close to both the $(\pi,\pi)$ nesting vector ($1.15~\text{\AA}^{-1}$) and the in-plane propagation vector in the parent compound CaFe$_2$As$_2$ ($1.14$~$\text{\AA}^{-1}$).~\cite{CaFe2As2}.
$Q_\text{res}$ is slightly larger than the $(\pi,\pi)$ nesting vector owing to the powder-averaging effect, as in the case of other polycrystalline iron-based superconductors.~\cite{1111_2,1111_3,111_1,Ba122_4,LiNH3Fe2Se2}

To elucidate the evolution of the spin resonance in CaKFe$_4$As$_4$, the temperature dependence of the integrated intensity of the spin resonance is plotted in Fig.~\ref{Fig:Q}(b).
To reduce the ambiguity of the fitting, the peak center and the peak width of the Gaussian function were fixed at their values at 4~K [the solid line in Fig.~\ref{Fig:Q}(a)], and the scaling factor and the background were fitted to each $Q$ dependence at $T\ge10$~K in the same manner as used in previous works~\cite{111_1,LiODFeSe_1,11_1}
The intensity of the spin resonance is clearly enhanced below $T_\text{c}$, providing us with direct evidence that the observed spin resonance couples to the superconducting order parameter.

Above $T_\text{c}$, paramagnetic excitation was observed in CaKFe$_4$As$_4$, as shown in Fig.~\ref{Fig:Map}(b).
The spectral weight of the spin resonance shifts to the lower energy [Figs.~\ref{Fig:Map}(b) and \ref{Fig:Energy}(a)].
As plotted in Fig.~\ref{Fig:Q}(a), the peak center of the paramagnetic excitation is close to $Q_\text{res}$, and the peak becomes broader because of the short correlation length.
Indeed, recent RPA calculations on the normal state in CaKFe$_4$As$_4$ suggested the existence of strong scattering at the $(\pi,\pi)$ nesting positions.~\cite{CaKFe4As4_DFT}
Similar paramagnetic excitation was also observed in the paramagnetic phase of superconducting Ba$_{0.6}$K$_{0.4}$Fe$_2$As$_2$,~\cite{Ba122_1} LaFeAsO$_{0.943}$F$_{0.057}$,~\cite{1111_4} and $\beta$-Fe$_{1+x}$Se.~\cite{11_1}

Our INS measurements on the dynamical spin susceptibility in CaKFe$_4$As$_4$ successfully captured two important features: 
(1) The observed spin resonance is localized at the $(\pi,\pi)$ nesting vector.
(2) The spin-resonance energy is clearly smaller than $<$$2\Delta$ ($=25$~meV) obtained from the previous ARPES measurements.~\cite{1144_ARPES}
These features provide strong evidence in favor of the sign-changing $s_\pm$-pairing symmetry in CaKFe$_4$As$_4$.~\cite{spm_3}
This result is further supported by our RPA calculations, as described below.

\begin{figure}[t]
\includegraphics[width=8.47cm]{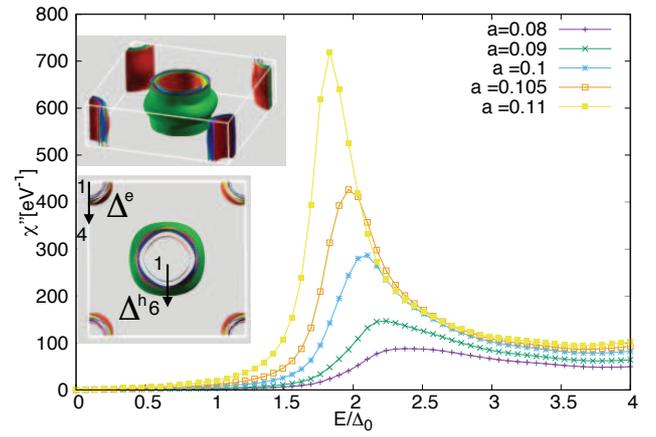}
\caption{\label{Fig:RPA}
(Color online)
Dynamical spin susceptibility $\chi''(E)$ of CaKFe$_4$As$_4$ obtained by using the multiorbital RPA with use of the 20-orbital 3D tight-binding model at $Q=(\pi,\pi,0)$.
We multiply all the electron--electron interactions by a factor $a$.
The insets show Fermi surfaces obtained by the 20-orbital 3D tight-binding model.
}
\end{figure}

We calculated the dynamical susceptibility on the basis of the multiorbital RPA~\cite{spm_2,Ba122_4} with the use of the effective 20-orbital three-dimensional (3D) tight-binding model at $T=0$.
For the effective model, maximally localized Wannier functions from the 20 Fe $3d$ bands were constructed.~\cite{Mostofi}
The electronic structure calculations were performed with the WIEN2K package,~\cite{Blaha} employing the full-potential linear augmented plane wave (FP-LAPW) method.~\cite{Koelling}
The generalized gradient approximation (GGA) to the exchange-correlation potential in the Perdew-Burke-Ernzerhof (PBE) form was used.
We confirmed that the resultant band structure and Fermi surfaces are consistent with those in previous work.~\cite{CaRbFe4As4_1}
There are six hole-like Fermi surfaces around the $\Gamma$ point and four electron-like Fermi surfaces around the $M$ point, as depicted in the insets of Fig.~\ref{Fig:RPA}.
We assumed the isotropic superconducting gaps $\Delta_{1}^{h}$, $\Delta_{2}^{h}$, $\Delta_{3}^{h}$, $\Delta_{4}^{h}$, $\Delta_{5}^{h}$, $\Delta_{6}^{h}$ and $\Delta_{1}^{e}$, $\Delta_{2}^{e}$, $\Delta_{3}^{e}$, $\Delta_{4}^{e}$ in  ascending order of the size of the hole-like and electron-like Fermi surfaces.~\cite{1144_ARPES}
According to the ARPES experiment,~\cite{1144_ARPES} we set $\Delta_{1}^{h}=(5/6)\Delta_{0}$, $\Delta_{2}^{h}=\Delta_{0}$, $\Delta_{3}^{h}=\Delta_{4}^{h}=\Delta_{5}^{h} = (11/12)\Delta_{0}$, $\Delta_{6}^{h}=(1/2)\Delta_{0}$, and $\Delta_{1}^{e}=\Delta_{2}^{e}=\Delta_{3}^{e}=\Delta_{4}^{e}=-\Delta_{0}$.
We employ the orbital-interaction coefficients $U_{s} = a U_{qt,M}^{rs}$ obtained by first-principles calculation for the 122 systems.~\cite{Miyake}
All the electron--electron interactions were multiplied by a factor $a$, since realistic values of the interaction results in very large spin fluctuations in the RPA calculation.
The number of $k$-meshes was $96 \times 96 \times 8$ and the smearing factor was $\eta = \Delta_{0}/8$.
We set $\Delta_{0} = 0.1$~eV to avoid numerical difficulty.

Figure \ref{Fig:RPA} shows the calculated dynamical spin susceptibility $\chi''(E)$ at $Q=(\pi,\pi,0)$.
The peak resulting from the resonance of the $s_{\pm}$-wave Cooper pairing is enhanced with increasing electron--electron interaction, which is similar to the results of the theoretical calculations with $s_{\pm}$-wave pairing in other iron-based superconductors \cite{Korshunov,NagaiKuroki}.
However, the peak structure of the present calculation is broad, since the superconducting gap value is different on each hole Fermi surface.
This RPA calculation can explain the experimental spectrum that is broad in energy, as shown in Fig.~\ref{Fig:Energy}(b).

In summary, we investigated the dynamical spin susceptibility in the new-structure-type iron-based superconductor CaKFe$_4$As$_4$.
The spin resonance at the $(\pi, \pi)$ nesting wave vector was enhanced below $T_\text{c}$.
$\chi''(\hbar\omega)$ of the spin resonance is broad, representing the different superconducting gaps on different portions of the Fermi surfaces.

This research at ORNL's Spallation Neutron Source was sponsored by the Scientific User Facilities Division, Office of Basic Energy Sciences, U.S. Department of Energy.
The calculations were performed on the supercomputing system SGI ICE X at the Japan Atomic Energy Agency.
Sample characterization was performed by using the SQUID magnetometer (MPMS, Quantum Design Inc.) at the CROSS user laboratories.
Travel expenses for the ARCS experiment were provided by the General User Program for Neutron Scattering Experiments, Institute for Solid State Physics, The University of Tokyo (Proposal Number GPTAS:16913), at JRR-3, Japan Atomic Energy Agency, Tokai, Japan.
The present work was partially supported by JSPS KAKENHI Grant Numbers JP15K00178, JP15K17686, JP15K17712, and JP17K14349.
This study was also supported by the ``Topological Materials Science'' (No. JP16H00995) KAKENHI on Innovative Areas from JSPS of Japan and the Cooperative Research Program of ``Network Joint Research Center for Materials and Devices'' (2015143 and 20161060).


\end{document}